\documentclass[twocolumn,showpacs,amsmath,amssymb,nofootinbib,pra,superscriptaddress]{revtex4-1}

\usepackage{comment}

\usepackage{xcolor}

\usepackage{graphicx}
\usepackage{epsfig}
\usepackage{dcolumn}
\usepackage{bm}
\usepackage{braket}
\usepackage{amsmath}
\usepackage{mathtools}
\usepackage{color}

\usepackage{multirow}

\begin{document}


\title{A Solvable Model for Decoupling of Interacting Clusters}

\author{Artem G. Volosniev}
\affiliation{Institut f{\"u}r Kernphysik, Technische Universit{\"a}t Darmstadt, 64289 Darmstadt, Germany} 
\author{Aksel S. Jensen}
\affiliation{Department of Physics and Astronomy, Aarhus University, DK-8000 Aarhus C, Denmark} 
\author{Nathan L. Harshman}
\affiliation{Department of Physics, American University, 4400 Massachusetts Ave. NW, Washington, DC 20016, USA}  
\author{Jeremy R. Armstrong}
\affiliation{Department of Physics and Astronomy, University of Nebraska at Kearney, 2401 11th Avenue, Kearney 68849, USA} 
\author{ Nikolaj T. Zinner}
\affiliation{Department of Physics and Astronomy, Aarhus University, DK-8000 Aarhus C, Denmark} 
\affiliation{Aarhus Institute of Advanced Studies, Aarhus University, DK-8000 Aarhus C, Denmark}

\date{\today}

\begin{abstract}
We consider $M$ clusters of interacting particles, whose
in-group interactions are arbitrary, and inter-group
interactions are approximated by oscillator potentials. We show
that there are masses and frequencies that decouple the in-group and
inter-group degrees of freedom, which reduces 
the initial problem to $M$ independent problems that describe each of 
the relative in-group systems. The dynamics of the $M$ center-of-mass coordinates
is described by the analytically solvable problem of $M$ coupled harmonic oscillators.
This letter derives and discusses these decoupling conditions. Furthermore, to illustrate our findings, we consider a charged impurity interacting with a ring of ions.
We argue that the impurity can be used to probe the center-of-mass dynamics of the ions.
\end{abstract}

\maketitle

\section{Introduction}

Exactly solvable problems play an important role in physics~\cite{suth2004}.  They are tools to check approximation schemes~\cite{lieb1963}, to learn about fundamental phenomena~\cite{onsager1944}, and to test numerical codes~\cite{white1994}. In addition, they give an incentive to find relevant realistic systems; see for example~\cite{johanning2009, bloch2012, guan2013} and references therein for recent cold-atom and cold-ion quantum simulators. In this paper we study (arguably) the simplest exactly solvable model --- a system of coupled harmonic oscillators. In spite of its simplicity, it plays a significant role in quantum mechanics where it describes the behavior of systems close to an equilibrium position~\cite{feynman1965}. Furthermore, coupled oscillators are a standard platform for model studies. For example, in few-body physics they are employed to estimate energies and gain insight into spatial correlations~\cite{kestner1962,mosch1996}. In many-body physics they provide a route for analyzing statistical phenomena such as quantum Brownian motion~\cite{ford1965} and quantum quenches~\cite{cardy2007,polk11}. 

A peculiar feature of harmonic coupling is that it can lead to solvable models even if other types of interactions are present in the system. In such models groups or clusters can be identified that have non-harmonic interaction within them while maintaining harmonic coupling between the clusters.  For example, the simplest Hooke's atoms and molecules are integrable because the in-group motion (e.g., ``electron-electron'' for atoms) decouples from the inter-group dynamics (``electron-nucleus'')~\cite{kestner1962,kais1993, pino1998, ugalde2005}. Surprisingly, this feature survives in  many-body systems that can be divided into clusters with not more than two particles per cluster. This was recently demonstrated in refs.~\cite{karwowski2008, karwowski2010, armstrong2015}. In this letter we show that the in-group and inter-group dynamics can be decoupled
also for systems with more than two particles per group. 
This decoupling provides a reference point for numerical and theoretical studies, and generates new separable and integrable models.  
Moreover, it leads to a solvable model for studying the appearance of clusters from microscopic Hamiltonians. Clustering is ubiquitous in nature: as we move up in scale, quarks and gluons combine into nucleons, nucleons bind into nuclei, nuclei and electrons bind into atoms, atoms into molecules and so on. At each stage, an effective description by bound clusters emerge. The ``natural" degrees of
freedom to understand the physics of the system are those of the
clusters, and the internal degrees of freedom of the constituent
particles lose relevance. Our paper provides a concrete, solvable
example of how this can occur. In this example, the in-group and
inter-group dynamics decouple, and the effects of tracing out the
constituent scale can be exactly evaluated.

\begin{figure}
\centerline{\includegraphics[scale=0.5]{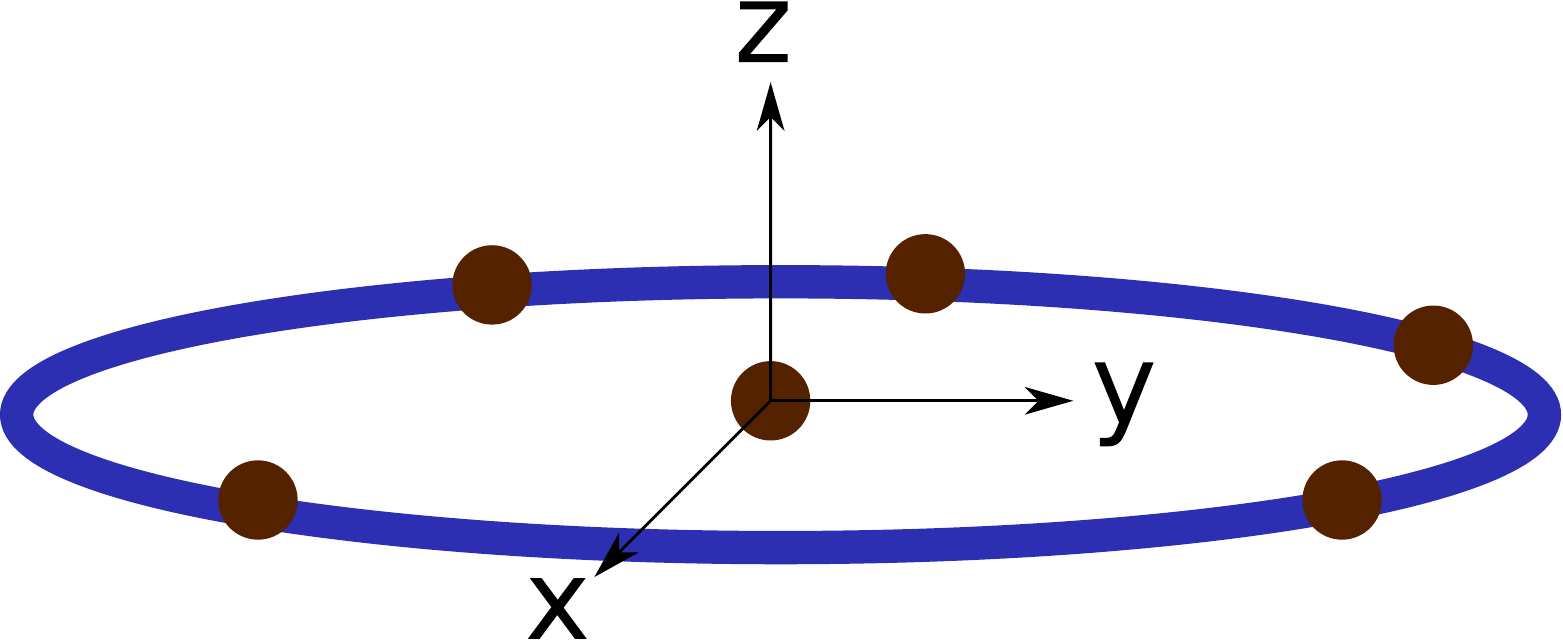}}
\caption{A sketch of the bath-impurity system, in which the impurity (in the middle) is coupled equally-strongly to all particles in the bath (in the ring). This geometry is inspired by the experiments with cold ions~\cite{li2017}.}
\label{fig:ill}
\end{figure}

\section{Model}
\label{sec:gen_model}

We consider $M$ clusters of particles with harmonic 
interactions between the clusters. 
The cluster $\alpha$ [$\alpha=1,2,3,\dots,M$] contains
$N_{\alpha}$ particles interacting via arbitrary pairwise
potentials, $V^{(\alpha)}_{ik}$ [$i,k=1,2,3, \dots, N_{\alpha}$]. 
In addition, the cluster may be subject to external one-body harmonic oscillator
potential, which for simplicity we take to be isotropic. The extension of our work to the non-isotropic case is straightforward; see our impurity-in-bath example. 
The particle $i$ in the cluster
$\alpha$ has the mass $m^{(\alpha)}_{i}$, and the coordinate
$\mathbf{r}^{(\alpha)}_{i}$.  The Hamiltonian, $H$, for the
system reads
\begin{eqnarray}
  H &=& \sum^{M}_{\alpha=1} H_{\alpha} +  
 \sum^{M}_{\alpha >\beta= 1} U_{\alpha \beta} \;,   \label{eq20} \\
  H_{\alpha} &=&  \sum^{N_{\alpha}}_{i=1} \left(
 -\frac{\hbar^2}{2 m^{(\alpha)}_{i}} \vec{\nabla}^2_{\mathbf{r}^{(\alpha)}_{i}}
 + \frac{m^{(\alpha)}_{i} \left(\omega^{(\alpha)} \mathbf{r}^{(\alpha)}_i\right)^2 }{2}  \right)  \nonumber \\
  &+& \sum^{N_\alpha}_{i > k = 1}  V^{(\alpha)}_{ik}\left(\mathbf{r}^{(\alpha)}_{i} - 
\mathbf{r}^{(\alpha)}_{k}\right) \; , \\ \label{eq25}
  U_{\alpha \beta} &=&  \frac{1}{2} \sum^{N_\alpha}_{i = 1} \sum^{N_\beta}_{k = 1}  
 d^{(\alpha \beta)}_{ik} \left(\mathbf{r}^{(\alpha)}_{i} -\mathbf{r}^{(\beta)}_{k}
  -  \mathbf{r}^{(\alpha \beta)}_{0} \right)^2,
\label{eq30}
\end{eqnarray}
where $d^{(\alpha \beta)}_{ik}$ characterizes the inter-group
harmonic interactions.  We allow for shifts of oscillator centers as expressed
by $\mathbf{r}^{(\alpha \beta)}_{0}$.  The one-body external potential is assumed to have a harmonic form, $m^{(\alpha)}_{i} (\omega^{(\alpha)} \mathbf{r}_i^{(\alpha)})^2/2$, and have one frequency, $\omega^{(\alpha)}$, for all particles in the same group, i.e., $\omega^{(\alpha)}$ does not depend on $i$.  These assumptions allow for exact separation of the in-group relative and center-of-mass coordinates. The two-body interactions, $V^{(\alpha)}_{ik}$, are left unspecified. The only assumption is that they only depend on the relative distances. Note that we could 
add a constant, $v^{(\alpha\beta)}_{ik}$, to each two-body oscillator  $U_{\alpha \beta}$.  The numerical calculations
would still be precisely the same, but this shift might be necessary to approximately reproduce the ground state energy of realistic systems \cite{arms11}.

\subsection{Decoupling} To show that in the Hamiltonian~(\ref{eq20}) the inter-group and in-group dynamics are decoupled we first notice that the couplings between the clusters $\alpha$ and $\beta$ are determined by the $ \mathbf{r}^{(\alpha)}_{i} \cdot \mathbf{r}^{(\beta)}_{k}$ terms:
\begin{equation}
U_{\alpha\beta}=w_{\alpha}(\mathbf{r}_i^\alpha)+w_{\beta}(\mathbf{r}_k^\beta)- \sum^{N_{\alpha}}_{i = 1} \sum^{N_{\beta}}_{k = 1} d^{(\alpha \beta)}_{ik} \mathbf{r}^{(\alpha)}_{i} \cdot \mathbf{r}^{(\beta)}_{k},
\label{eq:u_ab} 
\end{equation}
where the functions $w_{\alpha}(\mathbf{r}_i^\alpha)$ and $w_{\beta}(\mathbf{r}_k^\beta)$ are straightforwardly obtained from~eq.~(\ref{eq30}). In eq.~(\ref{eq:u_ab}) only the last term mixes the coordinates from different
clusters, $\alpha$ and $\beta$. If the couplings can be written as
\begin{eqnarray} \label{eq55}
  d^{(\alpha \beta)}_{ik} =   m^{(\alpha)}_{i}  m^{(\beta)}_{k} d^{(\alpha \beta)}_{0},
\end{eqnarray}
where $d^{(\alpha \beta)}_{0}$ is some parameter that does not depend on $i$ and $k$,
then the coupling term in eq.~(\ref{eq:u_ab}) reduces to
\begin{eqnarray} \label{eq60}
  \sum^{N_{\alpha}}_{i = 1} \sum^{N_{\beta}}_{k = 1} d^{(\alpha \beta)}_{ik}
  \mathbf{r}^{(\alpha)}_{i} \cdot \mathbf{r}^{(\beta)}_{k}  = 
 M_{\alpha}  M_{\beta} d^{(\alpha \beta)}_{0} 
 \mathbf{R}_{\alpha} \cdot \mathbf{R}_{\beta}  \; ,
\end{eqnarray}
where the total $\alpha$-group mass, $M_{\alpha}$, and the center-of-mass
coordinate, $\mathbf{R}_{\alpha}$, are defined by
\begin{eqnarray} \label{eq70}
 M_{\alpha} =     \sum^{N_{\alpha}}_{i = 1} m^{(\alpha)}_{i} \;\; ,\;\; 
 M_{\alpha} \mathbf{R}_{\alpha} =   \sum^{N_{\alpha}}_{i = 1} m^{(\alpha)}_{i}
 \mathbf{r}^{(\alpha)}_{i} \; .
\end{eqnarray}
Thus, with eq.~(\ref{eq55}) the couplings between the
clusters only involve their center-of-mass coordinates, 
which allows us to decouple the relative in-group motions and 
inter-group center-of-mass motions. We show in
the appendix that the assumptions in 
eqs.~(\ref{eq55}) are both necessary and sufficient for this decoupling (the exceptions for $N_{\alpha} \leq 2$ are discussed in ref.~\cite{armstrong2015}). 
 proceeding, we emphasize that in the most general case the coupling $d_{ik}^{\alpha\beta}$ must depend on the masses of particles, since the condition~(\ref{eq55}) requires the ratio $d_{ik}^{\alpha\beta}/(m_{i}^{\alpha}m_k^\beta)$ to be independent of the indexes $i$ and $k$. However, if the clusters $\alpha$ and $\beta$ are made of indistinguishable particles of type $A$ and $B$, correspondingly, then the condition~(\ref{eq55}) is automatically satisfied and $d_{ik}^{\alpha\beta}$ can be a mass-independent quantity; see the impurity-bath system on the next page.

With eqs.~(\ref{eq55}) and (\ref{eq60}) the Hamiltonian~(\ref{eq20}) can be rewritten as
\begin{eqnarray} \label{eq80}
 && H  = 
   H^{cm}  +   \sum^{M}_{\alpha=1} H_{\alpha}^{rel}  \; , \\ 
 && H_{\alpha}^{rel} =  - \sum^{N_{\alpha}}_{i = 1}
 \frac{\hbar^2}{2 m^{(\alpha)}_{i}} \vec{\nabla}^2_{\mathbf{r}^{(\alpha)}_{i}}  
 +  \sum^{N_\alpha}_{i > k = 1}  V^{(\alpha)}_{ik}\left(\mathbf{r}^{(\alpha)}_{i} 
   - \mathbf{r}^{(\alpha)}_{k}\right) \nonumber \\ \label{eq85}
  && + \frac{\hbar^2}{2 M_{\alpha}} \vec{\nabla}^2_{\mathbf{R}_{\alpha}}
 + \frac{1}{2}  \sum^{N_\alpha}_{i = 1} m^{(\alpha)}_{i} \left(\bar{\omega}^{(\alpha)}\right)^2 
  \left(\mathbf{r}^{(\alpha)}_{i} - \mathbf{R}_{\alpha}\right)^2 \; , \\ 
 &&  H^{cm}  = -\sum^{M}_{\alpha=1}\frac{\hbar^2}{2 M_{\alpha}} \vec{\nabla}^2_{\mathbf{R}_{\alpha}}
  +  \frac{1}{2}\sum^{M}_{\alpha=1} M_{\alpha} 
 \left(\omega^{(\alpha)} \mathbf{R}_{\alpha} \right)^2  \nonumber \\  \label{eq90}
 &+& \frac{1}{2}\sum^{M}_{\alpha>\beta=1} M_{\alpha} M_{\beta}
   d^{(\alpha \beta)}_{0} \left(\mathbf{R}_{\alpha} - \mathbf{R}_{\beta} - 
 \mathbf{r}^{(\alpha \beta)}_{0} \right)^2 \; ,
\end{eqnarray} 
where we defined the in-group  effective frequency arising from the
external one-body field and the inter-group coupling,
\begin{eqnarray} \label{eq95}
  \left(\bar{\omega}^{(\alpha)}\right)^2 \equiv \left(\omega^{(\alpha)}\right)^2 + 
   \sum^{M}_{\beta = 1 (\beta \neq \alpha)  } M_{\beta} d^{(\alpha \beta)}_{0} \; .
\end{eqnarray}
Note that in eq.~(\ref{eq85}) we subtracted the kinetic energy
operator for the group center-of-mass coordinate to maintain only the
relative coordinates in $H_{\alpha}^{rel}$.  To compensate, these
terms now appear in $H^{cm}$ where all center-of-mass dependencies are
collected. The operators $H^{cm}$ and $H^{rel}_\alpha$ commute,
 which leads to the decoupling of the relative in-group 
and inter-group center-of-mass motions.

\subsection{Spectrum} The observation that $[H^{cm},H^{rel}_\alpha]=[H^{rel}_{\beta},H^{rel}_\alpha]=0$ $\forall \alpha, \beta$ implies that the wave function and the energy that solve the Schr\"{o}dinger equation,  $H \Psi  = E\Psi$,
can be written as 
\begin{eqnarray} \label{eq110}
  \Psi &=& \Psi^{cm}(\{\mathbf{R}_{\alpha}\}) 
  \Pi_{\alpha=1}^{M} \Psi^{rel}_{\alpha}(\{\mathbf{r}^{(\alpha)}_{i}-\mathbf{R}_{\alpha}\}) 
 \; ,  \\  E &=&  E^{cm} + \sum^{M}_{\alpha=1} E^{rel}_{\alpha} \;, \label{eq112}
\end{eqnarray}
where the eigenvalue equations for $H^{cm}$ and $H^{rel}_\alpha$ read
\begin{eqnarray} \label{eq114}
 H^{cm} \Psi^{cm} = E^{cm} \Psi^{cm}  ,\;\;
 H_{\alpha}^{rel} \Psi^{rel}_{\alpha}  =  E^{rel}_{\alpha} \Psi^{rel}_{\alpha} \;.   
\end{eqnarray}
The Hamiltonian $H^{cm}$ describes a system of coupled oscillators, which is integrable for all values of the parameters~\cite{arms11} (see~\cite{gajda2000} for the most symmetric case: $M$ identical harmonically interacting particles). Therefore, to solve the initial problem with the $\sum_{\alpha}N_{\alpha}$ degrees of freedom we need to solve $M$ separate problems, each with $N_{\alpha}-1$ degrees of freedom.  

Note that if the spectra of the relative Hamiltonians $H^{rel}_{\alpha}$ can be calculated, then the same is true for $H$.  For example, if the potentials $V_{ik}^{\alpha}$ are zero range, i.e., $V_{ik}^{\alpha}(x)=g^{\alpha}_{ik}\delta(x)$, with every $g^{\alpha}_{ik}\to\infty$, then the spectrum of $H$ can be computed in the leading order in $1/g_{ik}^{\alpha}$~\cite{volosniev2014,volosniev2014a,deuret2014,levinsen2015}. Other relevant few-body systems are discussed in refs.~\cite{taut2003, werner2006}. Furthermore, if every $H^{rel}_{\alpha}$ corresponds to an integrable system (e.g., as in ref.~\cite{nathan2017}), then the decoupling implies that so is the Hamiltonian $H$.

\section{Impurity in a Bath} 
\label{sec:imp}

To illustrate the decoupling, we investigate the impurity-bath dynamics, where the bath is a system of identical particles in a ring, and the impurity is placed in the middle of the ring (see fig.~\ref{fig:ill}). Studies on impurity-bath dynamics shed light on thermalization mechanisms, and on the possibility to probe a bath employing impurities.  Coupled harmonic oscillators are one of the usual platforms for investigating these dynamics (see, e.g.,~\cite{ullersma1966,estrin1970,davies1973,Leggett1981,grabert1988,hanggi2005}), because, in particular, they can simulate a heat bath~\cite{ford1965}.  In those studies the impurity-bath coupling is often parametrized as $\mathbf{r}_I \sum c_i \mathbf{r}_i $, where $\{\mathbf{r}_i\}$ are the bath degrees of freedom, $\mathbf{r}_I$ is the coordinate of the impurity. The coupling coefficients $c_i$ are usually all different.  They are chosen to model realistic situations, for example, a chain of coupled harmonic oscillators with only nearest-neighbor interactions. We illustrate the cluster decoupling by considering the case $c_i=c_j$ $\forall i,j$, which can mimic the situation depicted in fig.~\ref{fig:ill}. In this situation the conditions in eq.~(\ref{eq55}) are satisfied, which means that the dynamics of the impurity is integrable. It is easy to understand the origin of the decoupling in this simple system. Indeed, the coupling is of the form $\sim \mathbf{r}_I \sum \mathbf{r}_i $, where $\sum \mathbf{r}_i$ defines the center-of-mass coordinate from eq.~(\ref{eq70}).

{\it Hamiltonian.} We consider an impurity particle in a bath, which consists of $N$ identical interacting particles. For simplicity, we assume that the impurity particle and every particle in the bath have the mass $m$.  All particles are confined by external linear oscillator potentials. The coordinate of the impurity is $\mathbf{r}_{I}=(x_I,y_I,z_I)$; the coordinate of the $i$th particle in the bath is $\mathbf{r}_{i}=(x_i,y_i,z_i)$. The Hamiltonian of the system is
\begin{equation}
H=H_I + H_B +  \sum_{i=1}^N \left(d_x x_ix_I+d_y y_iy_I+d_z z_iz_I\right),
\label{eq:eq_1}
\end{equation}
where the parameters $d_x,d_y$ and $d_z$  define the bath-probe coupling, and $H_I$, $H_B$ fully describe the probe and the bath when there is no coupling
\begin{align}
H_I&=-\frac{\hbar^2}{2 m} \vec{\nabla}^2_{\mathbf{r}_{I}} + \frac{m\omega_{x_I}^2 x_I^2}{2}+\frac{m\omega_{y_I}^2 y_I^2}{2}+\frac{m\omega_{z_I}^2 z_I^2}{2}, \\
H_B&=- \frac{\hbar^2}{2m}\sum_{i=1}^N \vec{\nabla}^2_{\mathbf{r}_{i}} +\sum_{i>j} V(\mathbf{r}_i-\mathbf{r}_j)
+\sum_{i=1}^N W(\mathbf{r}_i),
\end{align}
where $\{\omega_{x_I},\omega_{y_I},\omega_{z_I}\}$ are the trapping frequencies; $V(\mathbf{r})$ is the interaction potential and $W(\mathbf{r})$ is the one-body trapping potential for the particles in the bath.

{\it Motivation.} To motivate the study of the Hamiltonian~(\ref{eq:eq_1}), let us consider an ion placed in the middle of the ring of ions. This set-up is inspired by the current experiments with cold ions~\cite{li2017}. We assume that the impurity has a weaker confinement in the $z$-direction, such that the important degree of freedom is $z_I$. For the $i$th particle in the bath the confinement is such that the important degrees of freedom are $\phi_i, z_i$ ($\phi_i$ is the azimuthal angle). The corresponding Coulomb potential energy is written as 
\begin{equation}
k_{e}\frac{Q Q_I}{\rho\sqrt{1+\frac{(z_i-z_I)^2}{\rho^2}}}\simeq k_{e}\frac{Q Q_I}{\rho}\left(1-\frac{(z_i-z_I)^2}{2\rho^2}\right),
\label{eq:eq_4}
\end{equation}
where $\rho$ is the radius of the ring, $Q$ is the charge of a particle in the bath, $Q_I$ is the charge of the impurity, and $k_e$ is Coulomb's constant.  Here we assume that the relevant values of $z_i$ and $z_I$ are much smaller than $\rho$, i.e., $\langle z_i \rangle, \langle z_I \rangle \ll \rho$.  Therefore, the impurity-bath coupling is the same for all particles in the bath, and can be in the leading order described by eq.~(\ref{eq:eq_1}). Note that the assumption that the impurity is confined to the line piercing the middle of the ring is essential for our discussion: if the impurity could move in the $xy$-plane then eq.~(\ref{eq:eq_4}) would not be valid.

{\it Decoupling.} Motivated by the system of ions, we use in~eq.~(\ref{eq:eq_1}) $d_x=d_y=0$, and $\omega_{x_I}=\omega_{y_I}\to\infty$, the latter condition means that the impurity moves effectively in one spatial dimension described by $z_I$. Furthermore, we write $W(\mathbf{r_i})=v(x_i,y_i)+m\Omega^2z_i^2/2$, where $v(x_i,y_i)$ describes the ring trap in the $xy$-plane, and $\Omega$ is the trapping frequency in the $z$ direction. To show that the impurity degrees of freedom are decoupled from the relative motion of the particles in the bath, we introduce the Z-center-of-mass variable $N Z=\sum_i z_i$, which allows us to rewrite the Hamiltonian as 
\begin{equation}
H= H^{cm} + H^{rel},
\end{equation}
where $H^{cm}$ describes the coupling between the impurity and the center-of-mass coordinate
\begin{align}
H^{cm} &= -\frac{\hbar^2}{2m} \frac{\partial^2}{\partial z_I^2} -\frac{\hbar^2}{2m N}\frac{\partial^2}{\partial Z^2} \nonumber \\ &+ \frac{m \omega^2_{z_I} z_I^2}{2} + 
\frac{m N \Omega^2 Z^2}{2} + d_z N Zz_I;
\label{eq:HyR}
\end{align}
$H^{rel}$ describes the relative motion in the bath
\begin{align}
H^{rel} &= \sum_{i=1}^N \left(- \frac{\hbar^2}{2m}\vec{\nabla}^2_{\mathbf{r}_{i}}+v(x_i,y_i)+\frac{m\Omega\left(z_i-Z\right)^2}{2}\right) \nonumber \\&+\frac{\hbar^2}{2m N}\frac{\partial^2}{\partial Z^2} +\sum_{i>j} V(\mathbf{r}_i-\mathbf{r}_j).
\end{align}
The operators $H^{cm}$ and $H^{rel}$ depend on different variables, and, hence, they commute, i.e., $[H^{cm},H^{rel}]=0$.

The Hamiltonian $H^{cm}$ is solvable -- it describes a very-well studied system of two coupled harmonic oscillators~\cite{estes1968,scheid1987,kim1999, harshman2011, ueda2017, nagy2018}. Therefore, the decoupling allows one to study static and dynamic properties of the impurity in a simple manner.

{\it Applications of decoupling.} To illustrate the usefulness of the decoupling, we consider the quench dynamics in the system of ions from fig.~\ref{fig:ill} with characteristics similar to~\cite{li2017}, i.e., we use $^{40}$Ca$^+$ ions, $\rho=45\mu$m and $N=10$.  The corresponding Hamiltonian $H^{cm}$ is given by eq.~(\ref{eq:HyR})
with $d_z=k_e Q^2/\rho^3$ ($Q=Q_I$), $m \omega_{z_I}^2=m \omega_{ext}^2-N k_eQ^2/\rho^3, m \Omega^2=m \Omega^2_{ext}-k_e Q^2/\rho^3$, here $\omega_{ext}, \Omega_{ext}$ are the frequencies of the external potential. In what follows we assume that $\Omega=\omega_{z_I}\equiv \omega$. This assumption is not essential, but simplifies the theoretical calculations. It should also be noted that the radius of the ring is sufficiently large to prevent any tunneling between the impurity particle in the center and the ring particles, so that the $^{40}$Ca$^+$ ion in the center can function as an impurity despite being identical in nature to the bath ions.

We consider the following quench dynamics: At $t=0$ the system is in the ground state with $m \omega^2\gg d_z N$, which does not allow the impurity to couple to the bath. At $t>0$ the frequency $\omega$ is changed dynamically to $m \omega^2(t_f)\sim d_z N$, allowing for strong impurity-bath correlations. For the sake of argument, we use  $m \omega^2(t_f)=2d_z\sqrt{N}$.  The decoupling allows us to solve this seemingly complicated $1+N$ problem by investigating two coupled oscillators, and thus, relying on many previous studies. 

\begin{figure}

\includegraphics[scale=0.55]{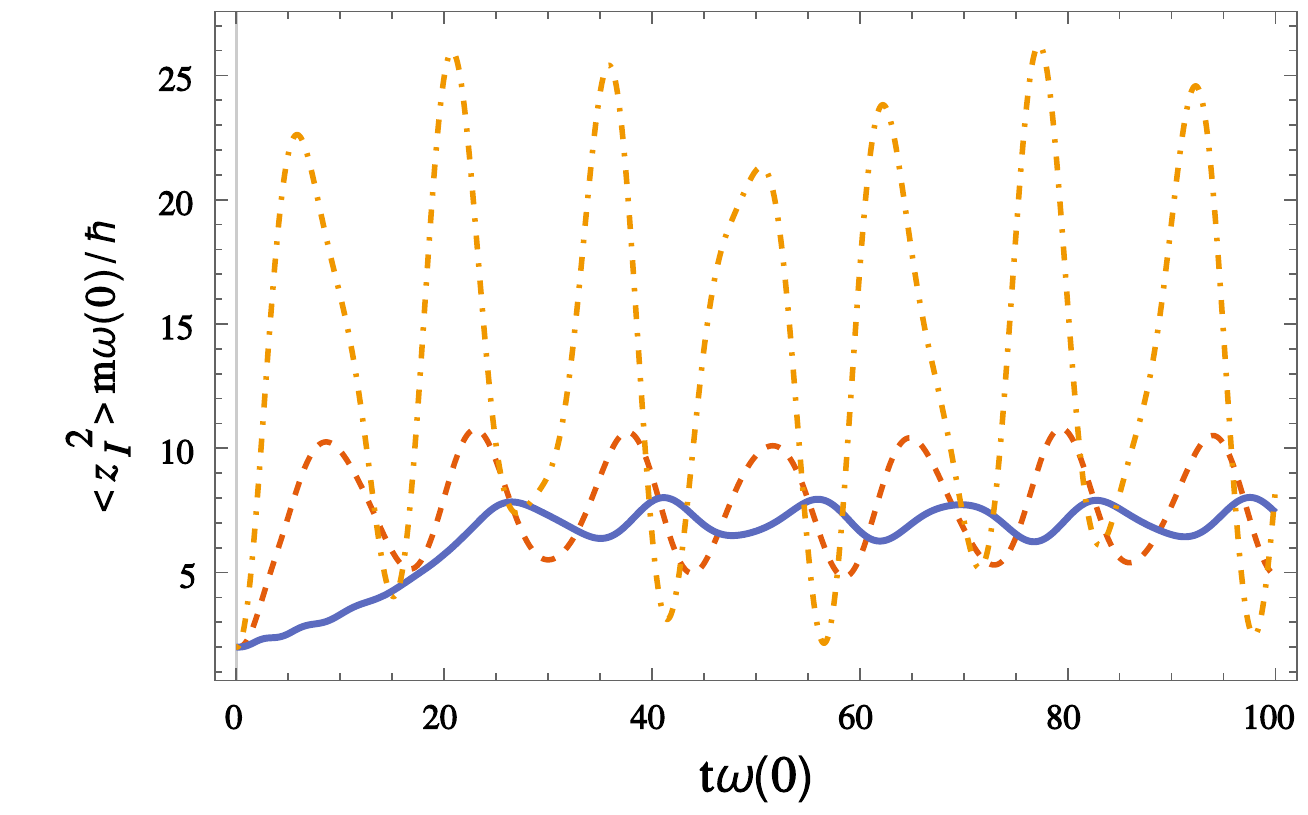}

\caption{The width of the impurity cloud $\langle z_I^2\rangle m\omega(0)/\hbar$ as a function of $t\omega(0)$ ($\omega(0)\simeq 1.5 \mathrm{MHz}$). Different curves show different drivings: the solid (blue) curve is for $\gamma=0.1$, the dashed (red) curve is for $\gamma=0.5$, and the dot-dashed (orange) curve is for $\gamma=10$.}
\label{fig:var}
\end{figure}

To find the solution, we introduce the variables
$x=(z_I-\sqrt{N}Z)/\sqrt{2}$ and $y=(z_I+\sqrt{N}Z)/\sqrt{2}$, in which the Hamiltonian $H^{cm}$ is written as 
\begin{equation}
H^{cm}=-\frac{\hbar^2}{2m} \frac{\partial^2}{\partial x^2} -\frac{\hbar^2}{2m}\frac{\partial^2}{\partial y^2}+ \frac{m \omega_x^2(t) x^2}{2}+\frac{m \omega_y^2(t) y^2}{2},
\end{equation}
where $m\omega_x^2(t)=m \omega^2(t)-d_z\sqrt{N}$, $m \omega_y^2(t)=m \omega^2(t)+d_z\sqrt{N}$. 
The solution to the Schr{\"o}dinger equation $H^{cm}\Phi=i\hbar\partial_t\Phi$ (assuming that the system is initially in the ground state) up to an irrelevant phase factor is~(cf.~\cite{husimi1953, popov1969,castin2004})
\begin{equation}
\Phi(x,y,t)=\psi_{\omega_x}(x,t)\psi_{\omega_y}(y,t),
\end{equation}
where 
\begin{equation}
\psi_{\omega_\xi}(\xi,t)=\frac{1}{\sqrt{\lambda_\xi}}\left(\frac{m \omega_\xi(0)}{\hbar\pi}\right)^{\frac{1}{4}}e^{\frac{m \xi^2}{2\hbar\lambda_\xi^2}\left(i\lambda_\xi \partial_t\lambda_\xi-\omega_\xi(0)\right)}
\end{equation}
with $\lambda^3_\xi(t)\partial_t^2\lambda_\xi(t)/\omega_{\xi}^2(0)=1-\omega_\xi^2(t)\lambda_\xi^4(t)/\omega^2_\xi(0)$; $\lambda_\xi(0)=1$ and $\partial_t\lambda_\xi(0)=0$.  The function $\Phi$ allows us to calculate all observables of interest. As an example, we calculate the variance $\langle z_I^2 \rangle (=N \langle Z^2 \rangle)$,
\begin{equation}
\langle z_I^2 \rangle= \frac{\int_{-\infty}^\infty |\psi_{\omega_x}|^2 x^2 \mathrm{d}x +\int_{-\infty}^\infty |\psi_{\omega_y}|^2 y^2 \mathrm{d}y}{2}.
\end{equation}
Note that $\int_{-\infty}^\infty |\psi_{\omega_\xi}|^2 \xi^2 \mathrm{d}\xi=\frac{\hbar\lambda_{\xi}^2(t)}{2 m\omega_\xi(0)}$, thus, the functions $\lambda_{x}$ and $\lambda_{y}$ alone determine 
$\langle z_I^2 \rangle$. We assume an exponentially fast transition from the initial to the final state: $m \omega^2(t)=20 d_z \sqrt{N}e^{-\gamma \omega(0) t}$ for $t<t_f$, and $m\omega^2(t)=20 d_z \sqrt{N}e^{-\gamma \omega(0) t_f}\equiv 2 d_z\sqrt{N}$ for $t>t_f$, where the dimensionless parameter $\gamma$ defines the speed of transition. 
This time dependence of $\omega(t)$ allows us to write the solution to the equation for $\lambda_\xi$ through Bessel functions (cf.~\cite{ebert2016}).  The corresponding variance $\langle z_I^2\rangle$ is plotted in fig.~\ref{fig:var}. For slow driving $\gamma=0.1$ we have an almost adiabatic evolution, whereas for faster driving $\gamma=0.5$ and $\gamma=10$ the evolution is more complex, but still readily calculable. Note that $\sqrt{\langle z_I^2\rangle}$ is much smaller than $\rho$ for all values of $\gamma$. Therefore, our approximation scheme for the Coulomb interaction is valid.

The dynamics of the system is easily computed using the Hamiltonian $H^{cm}$ also if the system is initially in the mixed state. This dynamics is relevant if the bath is in a finite temperature state at $t<0$. In this case, however, the choice of the initial state is not clear. 
Indeed, on the one hand, the center-of-mass of the bath is coupled only to the impurity by assumption. On the other hand, in reality there are higher-order terms in the potentials that will couple the $Z$-coordinate also to the relative motion in the bath. Since the dynamics depends strongly on the initial state, the impurity can be used as a probe to reconstruct the motional state of the $Z$-variable, similar to a sympathetic tomography~\cite{molmer2012}. The effect of a weak coupling to the environment might be studied using the master equation approach~(cf.~\cite{zhang1992}).

\section{Summary and Outlook}
\label{sec:concl}

In this article we consider clusters of interacting particles described by the Hamiltonian~(\ref{eq20}). Particles within clusters interact via arbitrary potentials, whereas the inter-group  interactions are of a harmonic form. We establish conditions~(\ref{eq55}) for which the Hamiltonian~(\ref{eq20}) can be written as a sum of commuting Hamiltonians: The intra-group and the inter-group dynamics are decoupled, which facilitates the description of such systems. 
For example, it simplifies calculations of certain observables such as inter-group entanglement, which will be focus of future research. Relevant studies have been 
performed in systems with all
harmonic interactions where entanglement measures can monitor the
decoupling for the Gaussian ground state and thermal states; see, e.g.,
refs.~\cite{plenio2002}. The strength of our model is that it is not 
constrained to only harmonic interactions.
Future directions for this work include quantifying the entanglement
that develops among clusters during the decoupling process, which should
still be an analytically tractable problem for our model.

More generally, whenever a system is partitioned into subsystems, either
theoretically or through some experimental control process, it is important to
understand how that external decoupling depends on the internal balance
of particle interactions. In the case of a partition into an impurity
probe and a bath, for example, what can we learn about the bath from the
dynamical decoupling of the probe? In this article we provide a specific
example of what information can be extracted from the bath by measuring
impurity probe observables.

We note that there are systems beyond what was presented in fig.~\ref{fig:ill} that can be studied with the harmonic approximation, and hence, the suggested model. In particular, the harmonic approximation is accurate if there is a well-defined minimum of the potential energy. Such a minimum often  occurs if particles interact via strong long-range potentials. A relevant textbook example is a crystal where the interactions are modelled by coupled oscillators. It is worthwhile noting that such a crystal while constructed of composite objects (ions, atoms, or molecules) can be readily described since the objects' center-of-mass motion is approximately decoupled from their internal degrees of freedom. Other prominent examples of systems with long-range interactions are atoms in cavities~\cite{long_range} and cold dipoles~\cite{cold_dipoles}. The harmonic approximation is useful for the former system if the sizes of atom traps are small in comparison to the length scales given by the cavity-mediated potentials. In the latter system, the harmonic approximation has been already successfully applied, in particular, to describe chains of cold dipoles 
in tubes and layers~\cite{volosniev2013, volosniev2013a,armstrong2012}. One can show, however, that the harmonic approximation obtained from a two-body dipole-dipole potential does not describe accurately more complicated structures, e.g., with more than one dipole per tube, and should be modified\cite{armstrong2019}. To this end, one might first calculate energies and structural properties of few-body structures (e.g., from~ref.~\cite{volosniev2013a}) and then use this knowledge to construct better harmonic models. 

In conclusion: We have presented a method to tremendously simplify $N$-body problems whose constituents can be divided into clusters with harmonic inter-group interactions.  We discuss a topical application of an impurity in a bath.

 \acknowledgements  We would like to thank Klaus M{\o}lmer for referring to~ref.~\cite{molmer2012}. This work has been supported by the Humboldt Foundation, the Deutsche Forschungsgemeinschaft (VO 2437/1-1) and the Ingenium organization of TU Darmstadt (A.~G.~V.); the Aarhus
University Research Foundation (N.~L.~H); the Danish Council for Independent Research and the DFF Sapere Aude program (N.~T.~Z.).

\section{Appendix}
\label{app:2}

The inter-group coupling is written in eq.~(\ref{eq:u_ab}) as
 $\sum^{N_{\alpha}}_{i = 1} \sum^{N_{\beta}}_{k = 1} d^{(\alpha \beta)}_{ik}
  \mathbf{r}^{(\alpha)}_{i} \cdot \mathbf{r}^{(\beta)}_{k}$.
The scalar product of the two vectors implies that the coupling is a
sum over the spatial dimensions of these vectors.  We can therefore
deal with each dimension separately, and subsequently add up the
contributions.  We first define the $x$-components of
$\mathbf{r}^{(\alpha)}_{i}$ collectively for $i=1,2,...,N_{\alpha}$ by
a new vector $\mathbf{x}^{(\alpha)}$.  The $x$-component of the coupling term 
is then written as
$U_{couple} =
  \mathbf{x}^{(\alpha){\dagger}}  
 \underline{D}^{(\alpha \beta)} \mathbf{x}^{(\beta)}$,
where the elements of the non-quadratic $N_{\alpha} \times N_{\beta}$
matrix $\underline{D}^{(\alpha \beta)}$ are $d^{(\alpha \beta)}_{ik}$.

We choose a new set of coordinates defined by $y^{(\alpha)}_i
= x^{(\alpha)}_i - {X}_{\alpha}$ for $i=1,2,...,N_{\alpha}-1$
and $y^{(\alpha)}_{i = N_{\alpha}} = {X}_{\alpha}$ (cf.~\cite{arms11}).  Here we
denote the $x$-coordinate of the center-of-mass vector
$\mathbf{R}_{\alpha}$ by ${X}_{\alpha}$.  This linear coordinate
transformation is described by a matrix, $\underline{T}^{(\alpha)}$, such that
$\mathbf{x}^{(\alpha)} =   \underline{T}^{(\alpha)} \mathbf{y}^{(\alpha)}$,
where the elements, $t^{(\alpha)}_{ik}$, of $\underline{T}^{(\alpha)}$ are:
$t^{(\alpha)}_{ik} = \delta_{ik}$ for $(i,k)=1,2,...,N_{\alpha}-1$;
$t^{(\alpha)}_{iN_{\alpha}}= 1$ for $i=1,2,...,N_{\alpha}$;
$t^{(\alpha)}_{N_{\alpha}k} = - m^{(\alpha)}_{k} / m^{(\alpha)}_{N_{\alpha}}$
for $k=1,2,...,N_{\alpha}-1$. The function $\delta_{ik}$ is the Kronecker delta.

The coupling potential $U_{couple}$ in the new coordinates reads
\begin{equation}\tag{A.2}  \label{eq50a}
 U_{couple} =
  \mathbf{y}^{(\alpha){\dagger}} \underline{D}^{(\alpha \beta)}_{T}
 \mathbf{y}^{(\beta)},
\end{equation}
where $\underline{D}^{(\alpha \beta)}_{T} =  \underline{T}^{(\alpha){\dagger}}
 \underline{D}^{(\alpha \beta)} \underline{T}^{(\beta)}$. Since 
all coordinates in $\mathbf{y}^{\alpha}$ and $\mathbf{y}^{\beta}$
are relative except the $y_{i = N_{\alpha}}^{(\alpha)}$ and $y_{k =
N_{\beta}}^{(\beta)}$ elements (i.e., ${X}_{\alpha}$ and ${X}_{\beta}$)
the cluster decoupling is achieved if and only if the transformation
leads to
\begin{equation}\tag{A.3}  \label{eq55a}
 U_{couple} =
  \mathbf{y}^{(\alpha)}_{N_{\alpha}} 
 \bigg(\underline{D}^{(\alpha \beta)}_{T}\bigg)_{N_{\alpha}N_{\beta}} 
 \mathbf{y}^{(\beta)}_{N_{\beta}}  \; 
\end{equation}
for all possible coordinates $\mathbf{y}^{(\alpha)}$ and $\mathbf{y}^{(\beta)}$.
This is achieved if and only if all matrix elements, except the last, 
are identically zero, that is 
$ \bigg(\underline{D}^{(\alpha \beta)}_{T}\bigg)_{ik} = 0$
for all $i$ and $k$ except when $(i,k) = (N_{\alpha},N_{\beta})$. 
These conditions can easily be worked out to give
\begin{align}  \label{eq70a}
\bigg(\underline{D}^{(\alpha \beta)}_{T}\bigg)_{ik} &= 
 d^{(\alpha \beta)}_{ik} 
  - \frac{m^{(\beta)}_{k}}{m^{(\beta)}_{N_{\beta}}} d^{(\alpha \beta)}_{i N_{\beta}} \nonumber \\ \tag{A.4} 
  &- \frac{m^{(\alpha)}_{i}}{m^{(\alpha)}_{N_{\alpha}}} d^{(\alpha \beta)}_{N_{\alpha} k} 
  + \frac{m^{(\alpha)}_{i} m^{(\beta)}_{k}}{m^{(\alpha)}_{N_{\alpha}} m^{(\beta)}_{N_{\beta}}} 
 d^{(\alpha \beta)}_{N_{\alpha} N_{\beta}},
\end{align}
for $i=1,2,...,N_{\alpha}-1$ and  $k=1,2,...,N_{\beta}-1$;
\begin{align}  \label{eq72a}
 \bigg(\underline{D}^{(\alpha \beta)}_{T}\bigg)_{N_{\alpha} k} = 
\sum_{i=1}^{N_{\alpha}}\left( d^{(\alpha \beta)}_{ik}   
- \frac{m^{(\beta)}_{k}}{m^{(\beta)}_{N_{\beta}}}  d^{(\alpha \beta)}_{iN_{\beta}} \right)  , \tag{A.5}
\end{align}
for $ k=1,2,...,N_{\beta}-1$;
\begin{align}  \label{eq74a}
\bigg(\underline{D}^{(\alpha \beta)}_{T}\bigg)_{i N_{\beta}} = 
 \sum_{k=1}^{N_{\beta}} \left(d^{(\alpha \beta)}_{ik}  
 - \frac{m^{(\alpha)}_{i}}{m^{(\alpha)}_{N_{\alpha}}} 
   d^{(\alpha \beta)}_{N_{\alpha} k}\right)   , \tag{A.6}
\end{align}
for $i=1,2,...,N_{\alpha}-1$; and 
\begin{align}  \label{eq76a}
 \bigg(\underline{D}^{(\alpha \beta)}_{T}\bigg)_{N_{\alpha} N_{\beta}} = 
 \sum_{i=1}^{N_{\alpha}} \sum_{k=1}^{N_{\beta}} d^{(\alpha \beta)}_{ik} \; . \tag{A.7}
\end{align}
Equations~(\ref{eq70a}), (\ref{eq74a}) and (\ref{eq72a}) give two identities
\begin{equation}\tag{A.8}  \label{eq78a}
  d^{(\alpha \beta)}_{i N_{\beta}} = \frac{m^{(\alpha)}_{i}}{m^{(\alpha)}_{N_{\alpha}}}
 d^{(\alpha \beta)}_{N_{\alpha} N_{\beta}} ,\;\;
 d^{(\alpha \beta)}_{N_{\alpha} k} = \frac{m^{(\beta)}_{k}}{m^{(\beta)}_{N_{\beta}}}
 d^{(\alpha \beta)}_{N_{\alpha} N_{\beta}} \;\;,
\end{equation}
for $i=1,2,...,N_{\alpha}-1$ and  $k=1,2,...,N_{\beta}-1$, 
which together with eq.~(\ref{eq70a}) give $
 d^{(\alpha \beta)}_{i k} = d^{(\alpha \beta)}_{N_{\alpha} N_{\beta}}
 \frac{m^{(\alpha)}_{i} m^{(\beta)}_{k}}{m^{(\alpha)}_{N_{\alpha}} m^{(\beta)}_{N_{\beta}}}$
for all $i$ and $k$. This necessary and sufficient condition is
equivalent to eq.~(\ref{eq55}). The remaining non-vanishing
matrix element in eq.~(\ref{eq76a}) is
\begin{align}  \label{eq82a}
 \bigg(\underline{D}^{(\alpha \beta)}_{T}\bigg)_{N_{\alpha}N_{\beta}} = 
 d^{(\alpha \beta)}_{N_{\alpha}N_{\beta}}
\frac{M_{\alpha}  M_{\beta}} {m^{(\alpha)}_{N_{\alpha}} m^{(\beta)}_{N_{\beta}}}.
\nonumber
\end{align}


\begin{thebibliography}{99}

\bibitem{suth2004} Bill Sutherland, {\it Beautiful Models: 70 Years of Exactly Solved Quantum Many-Body Problems},  World Scientific, River Edge, NJ, 2004.

\bibitem{lieb1963} One example is the test of Bogoliubov's perturbation theory using the exact solution to the Lieb-Liniger gas, see E. H. Lieb and W. Liniger, {\it Phys. Rev.} {\bf 130}, 1605 (1963).
\bibitem{onsager1944} One example is the study of critical phenomena using the exactly solvable two-dimensional square-lattice Ising model, see L. Onsager, {\it Phys. Rev.} {\bf 65}, 117 (1944).
\bibitem{white1994} One example is the test of the density matrix renormalization group using the exact solution to the antiferromagnetic  Heisenberg spin-1/2 chain, see S. R. White, {\it Phys. Rev. Lett.} {\bf 69}, 2863 (1994).
\bibitem{johanning2009} M. Johanning, A. F Var{\'o}n, and C. Wunderlich. {\it J. Phys. B: At. Mol. Opt. Phys.} {\bf 42}, 154009 (2009).
\bibitem{bloch2012} I. Bloch, J. Dalibard, and S. Nascimb{\'e}ne, {\it Nature Physics} {\bf 8}, 267 (2012).
\bibitem{guan2013} X.-W. Guan, M. T. Batchelor, and C. Lee,
{\it Rev. Mod. Phys.} {\bf 85}, 1633 (2013).
\bibitem{feynman1965} R. P. Feynman and A. R. Hibbs, {\it Quantum Mechanics
and Path Integrals}. McGraw-Hill, New York (1965).
\bibitem{kestner1962} N. R. Kestner and O. Sinano{\'g}lu, {\it Phys. Rev.} {\bf 128}, 2687 (1962).
\bibitem{mosch1996} M. Moshinsky and Y. Smirnov, {\it The Harmonic Oscillator in Modern Physics} (Amsterdam: Harwood Academic Publishers) (1996).
\bibitem{ford1965} G. W. Ford, M. Kac, and P. Mazur,{\it Journal of Mathematical Physics} {\bf 6}, 504 (1965).
\bibitem{cardy2007} P. Calabrese and J. Cardy, {\it J. Stat. Mech.} P06008 (2007).
\bibitem{polk11} A. Polkovnikov, K. Sengupta, A. Silva, and M. Vengalattore, \textit{Rev. Mod. Phys.} {\bf 83}, 863 (2011).

\bibitem{kais1993} S. Kais, D. R. Herschbach, N. C. Handy, C. W. Murray, and G. J. Laming, {\it J. of Chem. Phys.} {\bf 99}, 417 (1993).

\bibitem{pino1998} R. Pino and V. Mujica, {\it J. Phys. B: At. Mol. Opt. Phys.} {\bf 31}, 4537 (1998).

\bibitem{ugalde2005} E. V. Lude{\~n}a, X. Lopez, and J. M. Ugalde, {\it J. of Chem. Phys.} {\bf 123}, 024102 (2005).

\bibitem{karwowski2008} J. Karwowski, {\it J. Quantum Chem.} {\bf 108}, 2253 (2008).
\bibitem{karwowski2010} J. Karwowski and K. Szewc, {\it J.  Phys.:  Conf.  Ser.} {\bf 213}, 012016 (2010).
\bibitem{armstrong2015} J. R. Armstrong, A. G. Volosniev, D. V. Fedorov, A. S. Jensen, N. T. Zinner, {\it J. Phys. A: Math. Theor.} {\bf 48} 085301 (2015).



\bibitem{arms11} J. R. Armstrong, N. T. Zinner, D. V. Fedorov, and A. S. Jensen.  \textit{J. Phys. B} \textbf{44}, 055303 (2011).

\bibitem{gajda2000} M. A. Za\l{}uska-Kotur, M. Gajda, A. Or\l{}owski, and J. Mostowski, {\it Phys. Rev. A} {\bf 61}, 033613 (2000).

\bibitem{volosniev2014} A. G. Volosniev, D. V. Fedorov, A. S. Jensen, M. Valiente, and N. T. Zinner, \textit {Nature Commun.} {\bf 5}, 5300 (2014).

\bibitem{volosniev2014a} A. G. Volosniev, D. V. Fedorov, A. S. Jensen, N. T. Zinner, and M. Valiente,  \textit{Few-Body Syst} {\bf 55}, 839 (2014).

\bibitem{deuret2014} F. Deuretzbacher, D. Becker, J. Bjerlin, S. M. Reimann, L. Santos, {\it Phys. Rev. A} {\bf 90} 013611 (2014).

\bibitem{levinsen2015} J. Levinsen, P. Massignan, G. M. Bruun, M. M. Parish, {\it Science Advances} {\bf 1} e1500197 (2015).

\bibitem{taut2003} M. Taut, K. Pernal, J. Cioslowski, and V. Staemmler,
{\it J. of Chem. Phys.} {\bf 118}, 4861 (2003).

\bibitem{werner2006} F. Werner and Y. Castin, {\it Phys. Rev. Lett.} {\bf 97}, 150401 (2006).

\bibitem{nathan2017} N. L. Harshman, Maxim Olshanii, A. S. Dehkharghani, A. G. Volosniev, Steven Glenn Jackson, and N. T. Zinner, {\it Phys. Rev. X} {\bf 7}, 041001 (2017).
  

\bibitem{ullersma1966} P. Ullersma, {\it Physica}, {\bf 32}, 74 (1966).
\bibitem{estrin1970} Y. Z. {\'E}strin, {\it Radiophys Quantum Electron} {\bf 13}, 1474 (1970).
\bibitem{davies1973} E. Davies, {\it Commun. Math. Phys.}  {\bf 33}, 171 (1973).
\bibitem{Leggett1981} A. O. Caldeira and A. J. Leggett,
{\it Phys. Rev. Lett.} {\bf 46}, 211 (1981).
\bibitem{grabert1988} H. Grabert, P. Schramm, and G.-L. Ingold, {\it Physics Rep.}, {\bf 168}, 115 (1988).
\bibitem{hanggi2005} P. H{\"a}nggi and G.-L. Ingold, {\it Chaos} {\bf 15}, 026105 (2005).

\bibitem{li2017} H.-K. Li et al, {\it Phys. Rev. Lett.} {\bf 118}, 053001 (2017).

 
\bibitem{kim1999} D. Han, Y. S. Kim, and M. E. Noz, {\it American Journal of Physics} {\bf 67}, 61 (1998).
\bibitem{estes1968} L. E. Estes, T. H. Keil, and L. M. Narducci, {\it Phys. Rev.} {\bf 175}, 286, (1968).  
\bibitem{scheid1987} A. Sandulescu, H. Scutaru, and W. Scheid, {\it J. Phys. A: Math. Gen.} {\bf 20}, 2121 (1987).

\bibitem{harshman2011} N. L. Harshman and W. F. Flynn,  {\it Quantum Information and Computation}, {\bf 11} 278 (2011).

\bibitem{ueda2017} T. N. Ikeda, T. Mori, E. Kaminishi, and M. Ueda, {\it Phys. Rev. E} {\bf 95}, 022129 (2017).


\bibitem{nagy2018} I. Nagy, J. Pipek, and M.L. Glasser, {\it Few-Body Syst.}, {\bf 59}, 2 (2018).

 

\bibitem{husimi1953} K. Husimi, {\it Progress of Theoretical Physics} {\bf 9}(4), 381
(1953).
\bibitem{popov1969} V. S. Popov and A.M. Perelomov, {\it JETP} {\bf 29}, 738
(1969).
\bibitem{castin2004} Y. Castin, {\it Comptes Rendus Physique} {\bf 5}, 407 (2004).
\bibitem{ebert2016} M. Ebert, A. Volosniev, and H.-W. Hammer, \textit{Ann. Phys. (Berlin)}, {\bf 528}, 698 (2016).
\bibitem{molmer2012} S. Mirkhalaf and K. M{\o}lmer, {\it Phys. Rev. A} {\bf 85}, 042109 (2012).
\bibitem{zhang1992} B. L. Hu, J. P. Paz, and Y. Zhang, {\it Phys. Rev. D} {\bf 45} 2843, (1992).
\bibitem{plenio2002} K. Audenaert, J. Eisert, M. B. Plenio, and R. F. Werner, {\it Phys. Rev. A} {\bf 66}, 042327 (2002).
\bibitem{long_range} H. Ritsch, P. Domokos, F. Brennecke, T. Esslinger, Rev. Mod. Phys. {\bf 85}, 553 (2013).
\bibitem{cold_dipoles}  T. Lahaye, C. Menotti, L. Santos, M. Lewenstein, and T. Pfau, {\it Rep. Prog. Phys.} {\bf 72}, 126401 (2009).
\bibitem{volosniev2013} A. G. Volosniev, J. R. Armstrong, D. V. Fedorov, A. S. Jensen, N. T. Zinner, {\it Few-Body Systems} {\bf 54}, 707 (2013).
\bibitem{volosniev2013a} A. G. Volosniev, J. R. Armstrong, D. V. Fedorov, A. S. Jensen, M. Valiente, N. T. Zinner, {\it New J. Phys.} {\bf 15} 043046 (2013).

\bibitem{armstrong2012} J. R. Armstrong, N. T. Zinner, D. V. Fedorov, and A. S. Jensen, {\it Euro. Phys. J. D} {\bf 66}: 85 (2012).
\bibitem{armstrong2019} J. R. Armstrong, A. S. Jensen, A. G. Volosniev, and N. T. Zinner,  {\it Mathematics} {\bf 8}, 484 (2020).

\end{thebibliography}
\end{document}